\documentclass{elsart}

\newcommand{\be}{\begin{equation}}
\newcommand{\ee}{\end{equation}}

\newcommand{\ba}{\begin{eqnarray}}
\newcommand{\ea}{\end{eqnarray}}
\newcommand{\om}{\omega}
\newcommand{\Alfven}{ Alfv\'{e}n }

\begin{document}
\begin{frontmatter}
\title{ Role of Reconnection in AGN Jets}

\author{Maxim Lyutikov  $^{1,2,3}$}

\address{
$^1$ McGill University, 3600 University street, Montreal, QC, Canada H3A 2T8 \\
 $^2$ MIT, 77 Massachusetts Avenue, Cambridge, MA 02139\\
$^3$ CITA National Fellow}

\begin{abstract}
We discuss  the possible role of reconnection in  electro-magnetically dominated
cores of  relativistic AGN jets. 
We suggest that reconnection may proceed in a two-fold fashion:
initial explosive collapse on  the \Alfven  time-scale of a current-carrying jet
   (which is of the 
order of the  light crossing time)
and subsequent
slow quasi-steady reconnection. Sites of explosive collapse
are  associated with bright knots, while steady-state reconnection
re-energizes particles in the ``bridges'' between the knots.
 Ohmic dissipation in reconnection layers 
leads to particle  acceleration either  by inductive electric fields
or by stochastic particle acceleration in the ensuing electro-magnetic
turbulence.
\end{abstract}
\end{frontmatter}

Several arguments point to the possibility that in the  inner cores
of AGN jets  
the energy is transported mostly  by Poynting flux (see, e.g., Blandford 2002).    
 Perhaps the two strongest arguments in favor 
of magnetically dominated jets are the extremely strong collimation
of some jets (e.g., Pictor A) observed on a scale of tens of kiloparsecs
and a very short time-scale variability observed at TeV energies in blazars.
Unlike magnetically dominated jets, the hydrodynamic jets were expected
to be strongly influenced by interaction with the surrounding which should
lead to generation of oblique shocks, strong dissipation in the jets
and a loss of collimation. Apparently this does not happen, at least
in some jets. An alternative possibility 
is that jets consist  of a strongly magnetically  dominated
relativistic core, weakly relativistic sheath and later,
as the interaction with the external medium becomes important, 
a cocoon of the entrained material.

Recent observations of  TeV emission from blazars with
a very short time scale variability stressed once again the 
necessity for  {\it in situ} acceleration - the cyclotron decay times
for the X-ray and especially $\gamma$-ray emitting electrons are an order
of magnitude smaller than the light time travel from the core
(and often smaller than the light crossing time of the emitting region itself).
The two popular models of jet composition - leptonic and ionic -
both have difficulties in explaining the fast variability (a problem for 
ion-dominated jets, since cyclotron times are very long for ions)
 and total energy content 
(since leptons suffer strong radiative losses at the source due to radiation
drag). 

Magnetically  dominated jets are free from both of these constraints.
Magnetic energy is both ``quiet'' (is not subject to the compactness
problem and thus does not suffer from radiative losses at the source)
and of ``high quality''   - the 
energy stored in a low entropy
electro-magnetic outflows can be effectively converted into high 
frequency
electro-magnetic radiation {\it far away from the source}.
 Thus, the assumption of magnetically dominated jets implies that
the energy of the central source is released  in magnetic   form (e.g., by  the
Blandford-Znajek mechanism), 
transported as a Poynting flux  to large distances and then  dissipated
 by current instabilities.

Other  properties of magnetic fields may  argue
in favor of magnetically dominated jets.
(1) Magnetic fields are strongly non-linear systems: slow evolution
during which magnetic stresses build-up may lead to accumulation
of a large amount of  free energy which is  released on a short time
scale (of the order of the \Alfven  time scale) as the system crosses 
stability threshold. (2) Magnetic field
dissipation is usually strongly intermittent 
(as is exemplified by solar flares) and may be repetitive  in time 
 and/or in space.
(3) Magnetic field dissipation may be internal and does not necessarily lead
to the global disruption of a system.
(4) Magnetic field dissipation leads to particle acceleration
producing  power-law distribution  of accelerated
particles.

On the other hand, there are definite problems with reconnection models.
Reconnection physics is highly uncertain: it depends crucially on the
kinetic and geometric 
 properties of the plasma, which is very hard to test
observationally (e.g., Litvinenko 1999).
 Most importantly, various instabilities (based on 
inertial or cyclotron effects, 
 on ions or electrons) often seem to account  equally well
(with astrophysical accuracy) for the observed phenomena.
 This is an important uncertainty
since the principle  scale of the reconnection region (related, for example,
to  electron or  ion skin depth, Larmor radius or magnetic Debye radius)
is crucial in determining the rate of reconnection (e.g., Birk \& Lesch 2000). 
Thus, it is hard to produce a testable result that would give a decisive
answer to the question whether reconnection is important or not in the
AGN jets. This situation may be contrasted with the shock acceleration schemes,
where  a qualitatively correct result for the spectrum of accelerated particles
can be obtained from simple {\it macroscopic } considerations.

Currently, 
the most popular models of AGN jets assume that the dissipation is due to 
internal shocks related to time-depend source activity.
How different are the observational effects of reconnection from those of the internal
 shocks models?
Two separate questions are important here: that of particle acceleration
(spectra of particles, cut-off energies, etc) and emission
generation (synchrotron, IC, SSC, etc). 
Emission generation is likely to be the same in reconnection models, so that
most of the well detailed radiation models will still hold
(modulo, perhaps, the assumption of the  equipartition magnetic field).
On the other hand, the acceleration mechanism is very different, but 
  hardly distinguishable observationally -
both internal shocks
and reconnection regions represent  transient internal dissipative
regions, which heat the plasma and  accelerate particles. 
Spectra of accelerated particles are power-laws in both cases 
(power-law distributions are naturally produced   in 
shock acceleration; in the case of reconnection there is no  theoretical
expectation of power-laws, but they are observed both on the Sun and in
numerical experiments). A possible distinction between  shock and reconnection
acceleration is the value of the  power-law spectral index.
 Larrabee et al. (2002) have shown that in  relativistic reconnection
the spectrum of accelerated particles may be very hard,
$dn/d\gamma \sim \gamma^{-1}$; this is 
prohibitively hard for shock acceleration. In fact, such  {\it hard
power-laws  may be needed} to explain the TeV emission from blazars and
its propagation through the IR background.

We expect that  in  a relativistically moving  jet
the reconnection may
begin at distances $r \geq  R_S \Gamma^2 $ ($R_S$ is the Schwarzschild radius 
of the central black hole (BH)),
 which for a BH
of $10^8 \, M_{\odot}$ and $\Gamma \sim 30$ gives $ R \geq 10^{16}$ cm.
This is a fairly large distance, so  the compactness of the emitting
region is small and the high energy  photons are not likely to be scattered
inside the emitting region.

Reconnection is expected to develop in several stage.
(1) Slow (quiet) MHD evolution, build-up of magnetic stresses,
 $r \leq R_S \Gamma^2  \sim 10^{16}$ cm. 
(2) Explosive MHD collapse (sometimes called 
magnetic detonation or MHD catastrophe) developing 
 on an \Alfven time-scale and leading to formation
of current sheets. (3) Steady state reconnection.

{\bf Slow evolution}. 
Jets are launched in the vicinity of a BH-disk system.
In the process of acceleration they become super-fastmagnetosonic
and thus causally disconnected from the source.  
As the relativistic  flow expands it cannot adjust its  lateral balance 
on the expansion time scale due to 
effective freezing of lateral dynamics in relativistically moving jets.
This leads to the build-up of uncompensated stresses in the flow - a jet is in a
 strongly non-equilibrium state.

{\bf Explosive collapse}.
At distances $r \geq R_S \Gamma^2$  a jet starts to evolve laterally;
large uncompensated hoop stresses directed towards
the axis result in a jet compression. We propose that at this point the 
Poynting flux-dominated  core of a jet
 undergoes an electro-magnetic explosive collapse, similar to explosive
growth in  Z-pinches, high-$\beta$ TOKAMAKs disruptions,
geomagnetic substorms  and solar flares.
Alternatively, 
enhanced dissipation
 may be due to the onset of current instabilities that generate plasma
turbulence and result in anomalous resistivity.

{\bf Steady state reconnection}.
Explosive collapse
 drives a reconnection process.
The steady-state rates of reconnection
in AGN jets maybe estimated 
using the  formulation of  relativistic reconnection by
 Lyutikov \& Uzdensky (2002), who found that in  a strongly
magnetized plasma  the inflow velocity 
is $\beta_{\rm in} \sim \sqrt{ \sigma / S}$ (for $\sigma \leq S$),
where  $\sigma \gg 1 $ is the ratio of magnetic to plasma energy densities
and 
  $S$ is the Lundquist number $ S= { L c/\eta}  \gg 1$, $L$ is the size of the
reconnection layer and  $\eta$ is the resistivity. The outflowing plasma
is always relativistic, $\gamma_{\rm out} \sim \sigma \gg 1$.
Assuming that the dominant dissipation process is related to the ion
Larmor radius $r_L$ and using Bohm's prescription for resistivity,
$\eta \sim c r_L$ we find that typically the reconnection inflow velocities 
are non-relativistic,
$\beta_{\rm in} \sim  0.01 \left( {\sigma/{10^4}} \right)^{1/2} $.

Particle acceleration in reconnection regions may occur in a  variety of ways,
through DC acceleration, Fermi-type acceleration in
 MHD or electro-magnetic  turbulence and shock acceleration at the 
different kinds of MHD shocks produced near the reconnection region.
Simple qualitative estimates may be done assuming that particle
acceleration is due to DC electric fields and  is balanced by synchrotron
radiative losses. One then finds the {\it  maximum} Lorentz factor
and the {\it  maximum} energies for synchrotron and inverse Compton emission
\be
\gamma \sim \sqrt{ \beta_{\rm in}} \sqrt{ c  \over r_e \om_B},
\,\,\,\,
\epsilon_{\rm synch} \sim \beta_{\rm in} {  \hbar c \over r_e} \sim
20\,  \beta_{\rm in} \, {\rm MeV}
,
\,\,\,\,
\epsilon_{\rm IC}  \sim \gamma mc^2 \sim 100 {\rm TeV}
\ee
\vskip -.2 truein

where $r_e = e^2/mc^2$ and $\om_B = e B /mc$.
These values are able  to accommodate most observational constraints on the AGN 
emission properties.

Given all the possible advantages that reconnection may bring, 
can one "prove" that  it plays an important role in AGN jets?
Probably not 
 from  first principles. We expect that reconnection will become
an accepted model for jet re-energization either by some  particular
 analogy  to solar phenomena (e.g., coherent radio emission correlated 
with X-ray and $\gamma$-rays) or by exclusion 
(e.g., spectra prohibitively hard
for shock acceleration). Some of the observations may already be in place:
 some  knots have very  hard spectra   
(e.g., the  knot A1 of 3C 273 has
$\alpha \sim 0.6$, which requires an
electron acceleration spectrum with $p=1.2$, a challenge to any
shock acceleration model).

\end{document}